\documentclass[10pt,conference]{IEEEtran}
\IEEEoverridecommandlockouts
\usepackage{cite}
\usepackage{amsmath,amssymb,amsfonts}
\usepackage{algorithmic}
\usepackage{graphicx}
\usepackage{textcomp}
\usepackage{multirow}
\usepackage{xcolor}
\usepackage{booktabs} 
\usepackage{numprint}
\usepackage{caption}
\usepackage{subcaption}
\usepackage{listings}
\usepackage[export]{adjustbox}
\usepackage{xcolor}
\usepackage{hyperref}
\usepackage[T1]{fontenc}
\newcommand{\etal}{{\emph{et al. }}}
\usepackage{balance}
\newcommand{\eg}{{\emph{e.g.}}}
\definecolor{codegreen}{rgb}{0,0.6,0}
\definecolor{codegray}{rgb}{0.5,0.5,0.5}
\definecolor{codepurple}{rgb}{0.58,0,0.82}
\definecolor{backcolour}{rgb}{0.95,0.95,0.92}

\lstdefinestyle{mystyle}{
    backgroundcolor=\color{backcolour},   
    commentstyle=\color{codegreen},
    keywordstyle=\color{magenta},
    numberstyle=\tiny\color{codegray},
    stringstyle=\color{codepurple},
    basicstyle=\ttfamily\footnotesize,
    breakatwhitespace=false,         
    breaklines=true,                 
    captionpos=b,                    
    keepspaces=true,                 
    numbers=left,                    
    numbersep=5pt,                  
    showspaces=false,                
    showstringspaces=false,
    showtabs=false,                  
    tabsize=2
}

\def\BibTeX{{\rm B\kern-.05em{\sc i\kern-.025em b}\kern-.08em
    T\kern-.1667em\lower.7ex\hbox{E}\kern-.125emX}}
\begin{document}


\title{Large Language Models and Simple, Stupid Bugs
\thanks{This material is based upon work supported by the National Science Foundation under Grant NSF CCF (SHF-MEDIUM) No. 2107592. Any opinions, findings, and conclusions or recommendations expressed in this material are those of the author(s) and do not necessarily reflect the views of the National Science Foundation.}
}


\author{\IEEEauthorblockN{Kevin Jesse}
\IEEEauthorblockA{
\textit{UC Davis}\\
Davis, USA \\
krjesse@ucdavis.edu}
\and
\IEEEauthorblockN{Toufique Ahmed}
\IEEEauthorblockA{
\textit{UC Davis}\\
Davis, USA \\
tfahmed@ucdavis.edu
}
\and
\IEEEauthorblockN{Premkumar T. Devanbu}
\IEEEauthorblockA{
\textit{UC Davis}\\
Davis, USA \\
ptdevanbu@ucdavis.edu
}
\and
\IEEEauthorblockN{Emily Morgan}
\IEEEauthorblockA{
\textit{UC Davis}\\
Davis, USA \\
eimorgan@ucdavis.edu
}

}

\maketitle

\begin{abstract}
With the advent of powerful neural language models, 
AI-based systems to assist developers in  coding tasks are becoming
widely available; Copilot is one such system. 
Copilot uses Codex, a large language model (LLM), to complete code conditioned on a preceding ``prompt''. Codex, however, is trained on public GitHub repositories, \emph{viz.}, on code that may include bugs and vulnerabilities. Previous studies\cite{pearce2022examining, asare2022github} show Codex reproduces vulnerabilities seen in training. In this study, we examine how prone Codex is to generate an interesting bug category, single statement bugs, commonly referred to as simple, stupid bugs or SStuBs in the MSR community. We find that Codex  and similar LLMs do help avoid some SStuBs, but do produce \emph{known, verbatim} SStuBs as much as 2x as likely than \emph{known, verbatim} correct code. We explore the consequences of the Codex generated SStuBs and propose avoidance strategies that suggest the possibility of reducing the production of known, verbatim SStubs, and increase the possibility of producing known, verbatim fixes. 
\end{abstract}


\section{Introduction}
The rise of language-model based AI coding tools promises to change programming practice. Developers can now use AI coding tools, which inherit their power from models trained on enormous corpora of open-source code. Copilot, a language-model based coding assistant\cite{ziegler}, is available in many integrated development environments (IDEs). Copilot uses a model named Codex \cite{chen2021evaluating} to generate code completions. The full power of Codex is still being learned: it can already perform a diverse set of tasks including: code completion \cite{chen2021evaluating}, automatic program repair (APR) \cite{prenner2021automatic}, comment generation \cite{chen2021evaluating,fewshot2022ahmed}, program synthesis \cite{sobania2022choose, jain2022jigsaw}, and incident management \cite{ahmed2023recommending}. 

Copilot is free to use, and is widely adopted. It is an attractive tool for developers at different skill levels; it helps provide starting points for developers \cite{vaithilingam2022expectation} and can start functions from just input, output examples\cite{sobania2022choose}. The capabilities of Codex and similar models \cite{fried2022incoder,tunstall2022natural,xu2022systematic,nijkamp2022conversational} have raised many concerns, and have given rise to different research thrusts. Active avenues of Copilot research include how developers work with it: researchers report concerns on an over-reliance and unwarranted trust in Copilot-generated code \cite{sarkar2022like,vaithilingam2022expectation}, the quality of the completions \cite{yetistiren2022assessing,nguyen2022empirical}, security implications \cite{sandoval2022security,perry2022users,asare2022github, pearce2022asleep}, and copyright infringement \cite{karmakar2022codex}. These research topics are motivated by the free access and popularity of Copilot, particularly, the use of code it generates may give rise to broader ethical and functional concerns.

Codex has been found to work for program repair\cite{prenner2021automatic,pearce2021can}, problem solving\cite{karmakar2022codex,austin2021program}, math\cite{tang2021solving,drori2022neural}, and translation from natural language to various target languages\cite{chen2021evaluating, austin2021program, rajkumar2022evaluating, trummer2022codexdb} to name a few applications. With many use cases, Codex is a double-edged sword of utility and risk and ultimately we should find ways to minimize the risk and maximize the utility of Codex.

To that objective, this work examines Codex on the 2021 MSR mining challenge dataset ManySStubs4J\cite{karampatsis2020often}. The dataset consists of single statement `simple, stupid bugs' (SStuBs) mined from Maven projects. Karampatis and Sutton found that SStuBs have a frequency of 1 in every 1,600 LOC and that static analyzers cannot detect them\cite{karampatsis2020often,mosolygo2021rise}. Mosolyg\'{o} \etal \cite{mosolygo2021rise} determined that SStuBs appear more in larger chunks of code  authored by the same developer, perhaps due to loss of attention or misunderstanding of code functionality. We see this exact phenomena in Codex studies\cite{vaithilingam2022expectation, sarkar2022like} where developers use Codex to generate large blocks of code and, \textit{if} a bug is found, dive into a time-consuming rabbit hole to fix the code\cite{vaithilingam2022expectation}. Worse, this study reports that developers often blindly trust generated code, or optimistically hope to fix problems later. 

Surprisingly, so-called ``simple, stupid'' bugs can survive a long while; in the SStuBs dataset, fixes take around 240 days\cite{mosolygo2021rise,latendresse2021effective}. More worryingly, when Codex generates code, an `agent' other than the active IDE user is actually `coding', and thus the number of commits to fix the SStuB might be even longer (see Mosolyg\'{o} \etal \cite{mosolygo2021rise}). This disappointing possibility is supported by surveys\cite{vaithilingam2022expectation, sarkar2022like}, suggesting that  developers  don't always understand generated code, and struggle to fix any bugs therein.

The performance of Codex has been extensively benchmarked~\cite{chen2021evaluating, pearce2021can}, checked for security vulnerabilities\cite{pearce2022asleep, sandoval2022security,sandoval2022security,karmakar2022codex,perry2022users}, and empirically evaluated\cite{nguyen2022empirical,vaithilingam2022expectation,sarkar2022like,asare2022github}. However, Codex has not been evaluated against SStuBs, which are a special kind of bug~\cite{hua2021effectiveness}. To understand SStuBs related to AI-supported programming, we evaluate whether Codex and other code completion models produce SStuBs,  or their fixes; also 
we look at the consequences of such bugs in code bases. Finally, we present a Codex experiment aimed at \emph{automatically} 
communicating developer \textit{intent}, using \emph{generated} comments, to help avoid introducing simple, stupid bugs, and also producing commented code. 

Our research questions are as follows, all primarily evaluated using the ManySStuBs4J dataset:

\begin{itemize}
    \item \textbf{RQ1:} How often do Codex and similar language models (CodeGen{\cite{nijkamp2022conversational}} and PolyCoder~\cite{xu2022systematic}) produce simple, stupid bugs?
    \item \textbf{RQ2:} When Codex generates the same simple, stupid bug that a human does, how much time does the SStuB originally take to fix?
    \item \textbf{RQ3:} Does Codex produce buggy or correct code more confidently (\emph{viz.} at higher probability)?
    \item \textbf{RQ4:} Does adding automatically generated comments to the prompt help Codex and akin language models avoid SStuBs? Do other types of prompt improvements help reduce SStuBs? 
    \item \textbf{RQ5:} How do bug-derived code comments, when inserted in the prompt, affect code generated by Codex and other LLMs? 
\end{itemize}


Our key findings are: (1) LLMs do help avoid some SStuBS in our dataset! (2) Codex and other LLMs do produce \emph{known} SStuBs, and at a rather high rate (perhaps twice as often as they produce \emph{known} correct, bug-fixing code); (3) When Codex generates a known SStuB, it's associated (historically) with longer fix-times; (4) Codex-generated completions appear equally `natural'\cite{hindle2016naturalness}, regardless of whether they match buggy code or the related fix; if they match neither, they are less natural. (5) Automatically generated comments, when added to prompts, appear to reduce the known SStuB production rate for most models and improve the bug/patch ratio; (6) Even buggy comments help to reduce the bug/patch ratio in Codex, suggesting just attempting to comment code helps. The improvement in avoiding SStuBs with comments from neural comment generation model CodeTrans\cite{elnaggar2021codetrans}, suggest that using these models \textit{with} Codex would be beneficial to avoid SStuBs. 

Data from this study is available here\footnote{\url{https://doi.org/10.5281/zenodo.7676325}}.


\section{Related Work}
We discuss related work on language models and code quality. 

\subsection{Simple, Stupid Bugs}
Simple, stupid bugs (SStuBs) are bugs that have single-statement fixes that match a small set of bug templates. They are called ``simple'' because they are usually fixed by small changes and ``stupid'' because, \emph{once located}, a developer can usually  fix them quickly with minor changes. However, locating SStuBs can be time-consuming~\cite{karampatsis2020often}. Karampatsis  and Sutton \cite{karampatsis2020often} published a collection of SStuBs mined from a set of template bug types, \eg, \texttt{\small CHANGE\_IDENTIFIER} or \texttt{\small DIFFERENT\_METHOD\_SAME\_ARGUMENTS}. Through their study of the dataset, Karampatsis found that SStuBs are prevalent in code bases, accounting for 33\% of single statement bugs detected in 1000 Maven projects. They found that these bugs occur every 1,600 lines of code, and were not detected by static analysis. MSR once used ManySStuBs4J dataset as a mining challenge, to study these bugs, and manage their impact.

Mosolyg\'{o} \etal \cite{mosolygo2021rise} studied the history of SStuBs. They find that SStuBs are more frequent in code modified by the same developer, often when s/he writes large chunks of code. This is perhaps because such large coding tasks strain focus and attention.
They found that only 40\% of SStuBs were fixed by the same author in a median time of 4 days; when the SStuB is fixed by a different author, the SStuB took 136 days to find and fix!
We hypothesize that if comments were present in a Codex prompt, we'd get
better code completions, \emph{and} the entire chunk would be easier to read \& fix. 

Zhu and Godfrey \cite{zhu2021mea} studied how developers fix SStuBs. Similar to Mosolyg\'{o} \etal \cite{mosolygo2021rise} they found that developers fix their own bugs quicker, whereas bugs from other developers take significantly more time to fix. This suggests that Codex generated SStuBs \emph{may} take longer to fix since they come from an artificial `developer'.  Codex-generated code-snippets that include SStuBs may require extensively debugging to be patched in a similar fashion. A Copilot study \cite{vaithilingam2022expectation}, found that users had trouble debugging generated code from Codex spending considerable time and effort to fix, for instance, a generated regular expression. 

Madeiral and Durieux \cite{madeiral2021large} discussed SStuBs in the context of code clones\cite{roy2007survey} and the changes that introduce them, \emph{viz.} ``change clones''. They found that 29\% of change clones introduced SStuBs by matching the 16 SStuB patterns. Since Codex is a language model that tends to repeat code it's seen, it could conceivably generate SStuBs in multiple locations, increasing the repair effort.

Peruma and Newman \cite{peruma2021distribution} examined SStuBs in unit test files. They found that SStuBs tend to occur in non-test files and that developers fix the bugs separately despite test and non-test files being functionally related. Peruma and Newman also discovered that developers prioritize non-test files and the fixes in tests are associated with asserts. 

Latendresse \etal \cite{latendresse2021effective} and Hua \etal \cite{hua2021effectiveness} addressed the detection of SStuBs. Latendresse \etal \cite{latendresse2021effective} found that continuous integration (CI) tools cannot catch any SStuBs. Hua \etal \cite{hua2021effectiveness} found that deep learning vulnerability detectors were suboptimal compared to traditional vulnerability detectors on SStuBs. Our results confirm that models as large as Codex, find SStuBs to be equally regular to the patches it generates.

Mashhadi \etal \cite{mashhadi2021applying} applied CodeBERT\cite{feng2020codebert} (fine-tuned for patching) to SStuBs and could fix  19\% of de-duplicated Many4SStuBs4J dataset. Mashhadi \etal mentions an advantage of using CodeBERT: no special tokens are required like in SequenceR \cite{chen2019sequencer}; similarly, many APR techniques with Codex \cite{prenner2021automatic,pearce2021can} rely on a prior of knowing a bug exists or even, more specifically, the bug location \cite{chen2019sequencer}. Adding comments to the prompt does not require making any of these assumptions.

Finally, PySStuBs\cite{kamienski2021pysstubs} is a Python simple, stupid bug dataset. The more recent TSSB-3M\cite{richter2022tssb} is also a Python SStuBs dataset mined at scale. Our study focuses on the established Java SStuBs patterns from the everpopular ManySStuBs4J dataset. We plan to expand our findings to other languages and SStuB patterns, resources permitting. Currently OpenAI restricts usage of Codex to 20 requests per minute. Inference on large models for ManySStuBs4J takes just over a day.

\subsection{Examining Codex Completions}
Vaithilingam \etal \cite{vaithilingam2022expectation} studied the developer experience with Codex. The key findings were that most participants preferred Codex to Intellisense in Visual Studio IDE. Participants preferred to use Codex as a starting point in lieu of searching online. Unfortunately participants over-relied on Codex and then  struggled when generated code was buggy. The authors reported three major issues: (1) participants often didn't understand and assess the correctness of generated code, (2) participants underestimated the repair-effort required when generated code was buggy, (3) the prompts used by participants were quite varied, sometimes resulting in undesired code completions. 

Sarkar \etal \cite{sarkar2022like} wrote an extensive review of programming with an AI assistant Codex. Sarkar \etal surveys previous work citing Codex's reliability, safety, and security implications. The review covers studies in Codex usability, design, and user reports. Sandoval\cite{sandoval2022security} and Perry\cite{perry2022users} examine Codex security implications.

Yetistiren \etal \cite{yetistiren2022assessing} and Nguyen \etal \cite{nguyen2022empirical} empirically studied Copilot's code suggestions. Yetistiren \etal \cite{yetistiren2022assessing} found Copilot mostly generated valid code and Copilot improved it's correctness with further input from the developer; sample examples, unit tests, docstrings, and prompts increased correctness further. Nguyen \etal \cite{nguyen2022empirical} found Copilot correctness varies by programming language and does not differ in complexity (cognitive and cyclomatic) among programming languages.

Prenner \etal \cite{prenner2021automatic}, Pearce \etal \cite{pearce2021can,pearce2022examining}, Karmakar \etal \cite{karmakar2022codex}, and Ahmed \etal \cite{fewshot2022ahmed, ahmed2023recommending} apply Codex in various settings: automatic program repair (APR) \cite{prenner2021automatic}, security vulnerability prediction\cite{pearce2021can}, HackerRank challenges\cite{karmakar2022codex}, code summarization\cite{fewshot2022ahmed}, and incident management\cite{ahmed2023recommending}. In APR, Prenner \etal tried engineering prompts to find a way to push Codex to generate a non-buggy version of the code. Codex performed competitively to recent work but was sensitive in the prompt. Pearce \etal found Codex could repair 58\% of real world security vulnerabilities. Karmaker \etal applied Codex on HackerRank problems with great success; some of the success was attributed to Codex already knowing the solution despite an incomplete prompt, in other words, memorizing the solution. Ahmed \etal trains Codex on few-shot project-specific code to achieve state-of-the-art code summarization. Ahmed \etal found success in using Codex to help engineers diagnose and mitigate production incidents. All of these works use prompting to illicit a desired response from Codex.

Prompting instructions with natural language or code is often implemented as a comment to the prompt  passed to Codex. The prompts are implemented as comments to improve generated code, \eg, natural language instructions in 
docstrings or input-output examples. Prompting relies on knowledge \emph{a priori} that Codex should adopt into its generations. A prompt could be: an example completion, a problem description, input format, code that is vulnerable,  a docstring documenting a bug location, input-output pairs, and snippets of bugs and corresponding patches (few-shot). In this work, we explore similar prompting techniques from Prenner \cite{prenner2021automatic}, \textit{and} focus on \textit{prior-free} prompting through traditional commenting practices. We find that good commenting practices can guide Codex to more SStuB-free completions.





\begin{figure}
\centering

\includegraphics[width=\linewidth]{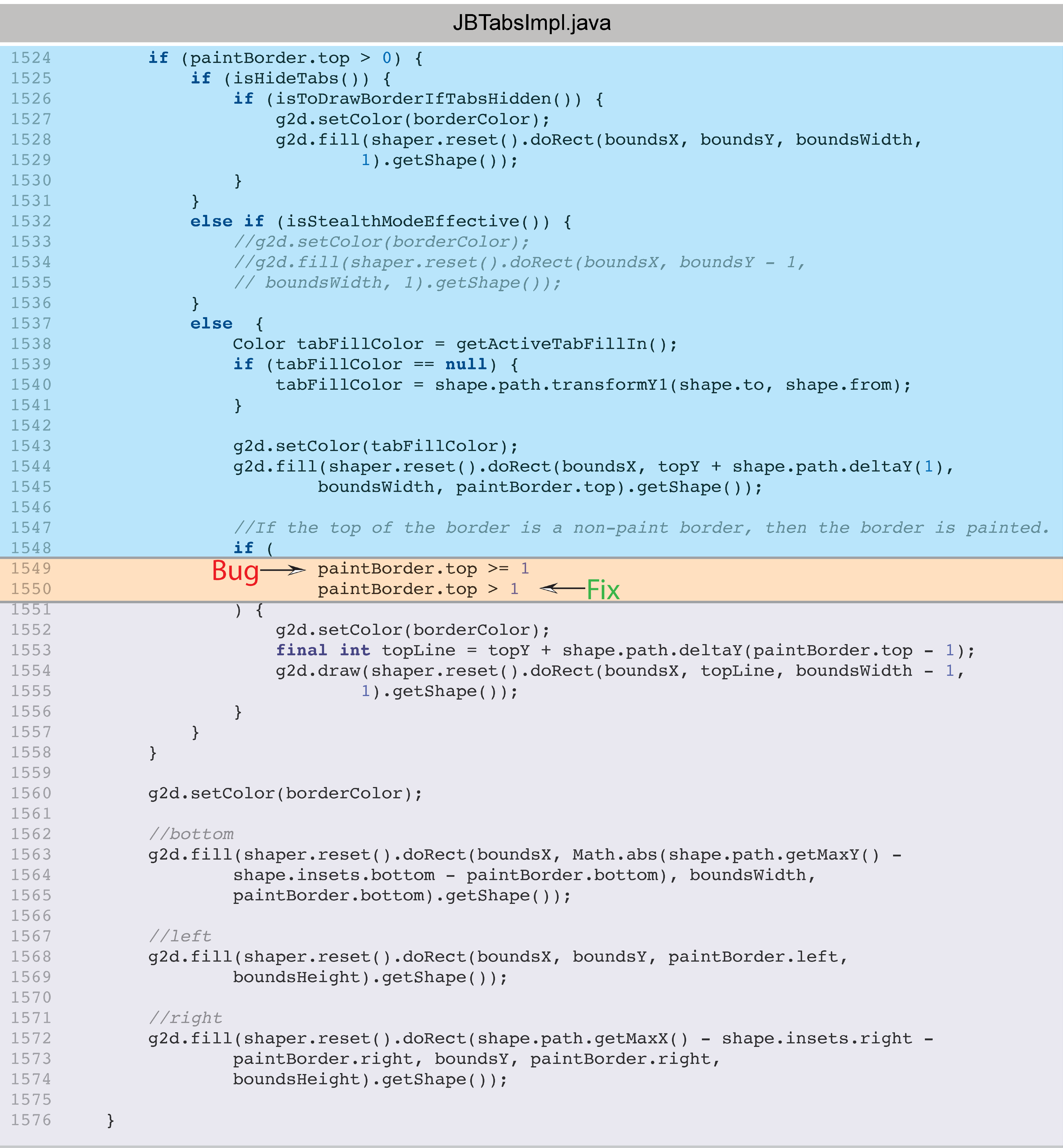}
  \caption{The orange highlighted code is the candidate single line completion that Codex can match to the automatic evaluation either the known bug or fix. Blue highlighted code is the prompt \textit{a.k.a.} the text proceeding SStuB statement that Codex uses for completion. The purple highlighted code is the code after the SStuB.}
  \label{fig:snippet1comment}

\end{figure}

\begin{figure}
\centering
\begin{subfigure}{.5\textwidth}
  \centering
  \includegraphics[width=\linewidth]{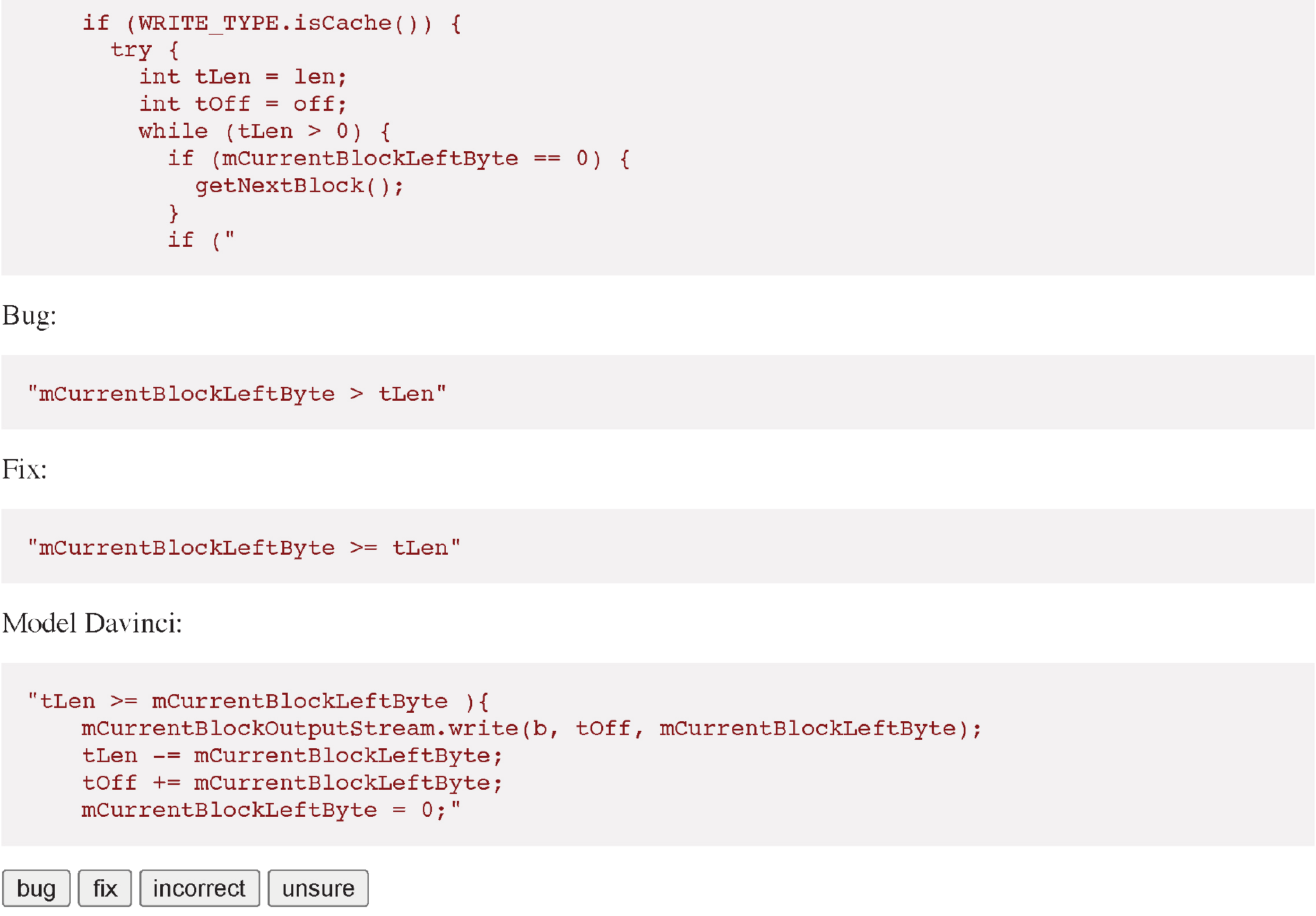}
  \caption{This completion is incorrect.}
  \label{fig:survey:incorrect}
\end{subfigure}%

\begin{subfigure}{.5\textwidth}
  \centering
  \includegraphics[width=\linewidth]{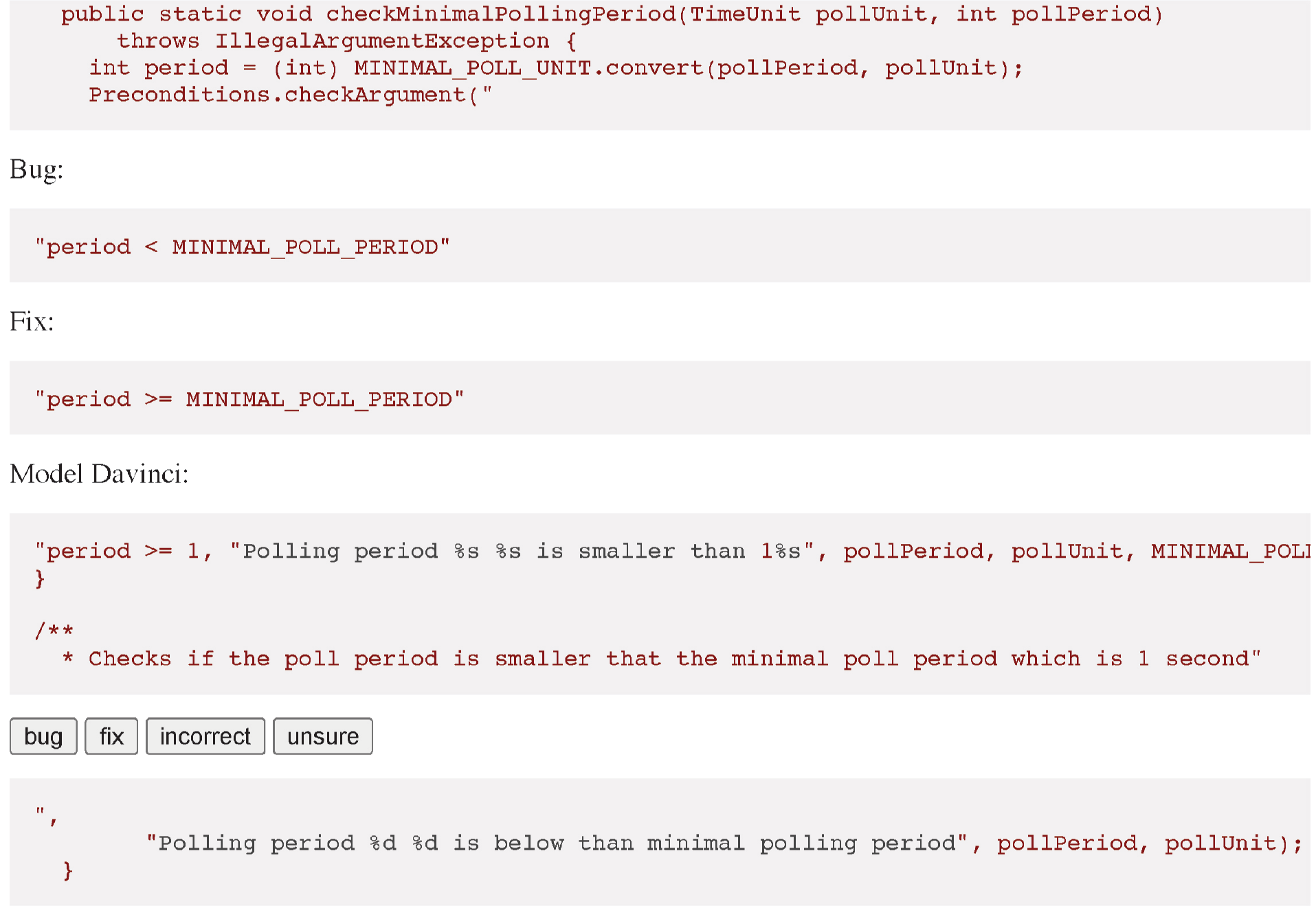}
  \caption{This completion matches the fix. Earlier in the code, the constant \texttt{\small MINIMAL\_POLL\_PERIOD} is set to 1.}
  \label{fig:survey:correct}
\end{subfigure}
\caption{This annotation tool helps mark Codex completions that do not match any SStuB directly. This guarantees our evaluation is not missing reasonable alternatives to the SStuB that could be deemed a bug or fix.}
\label{fig:survey}
\end{figure}

\begin{figure}[t]
    \centering
\includegraphics[width=.95\linewidth]{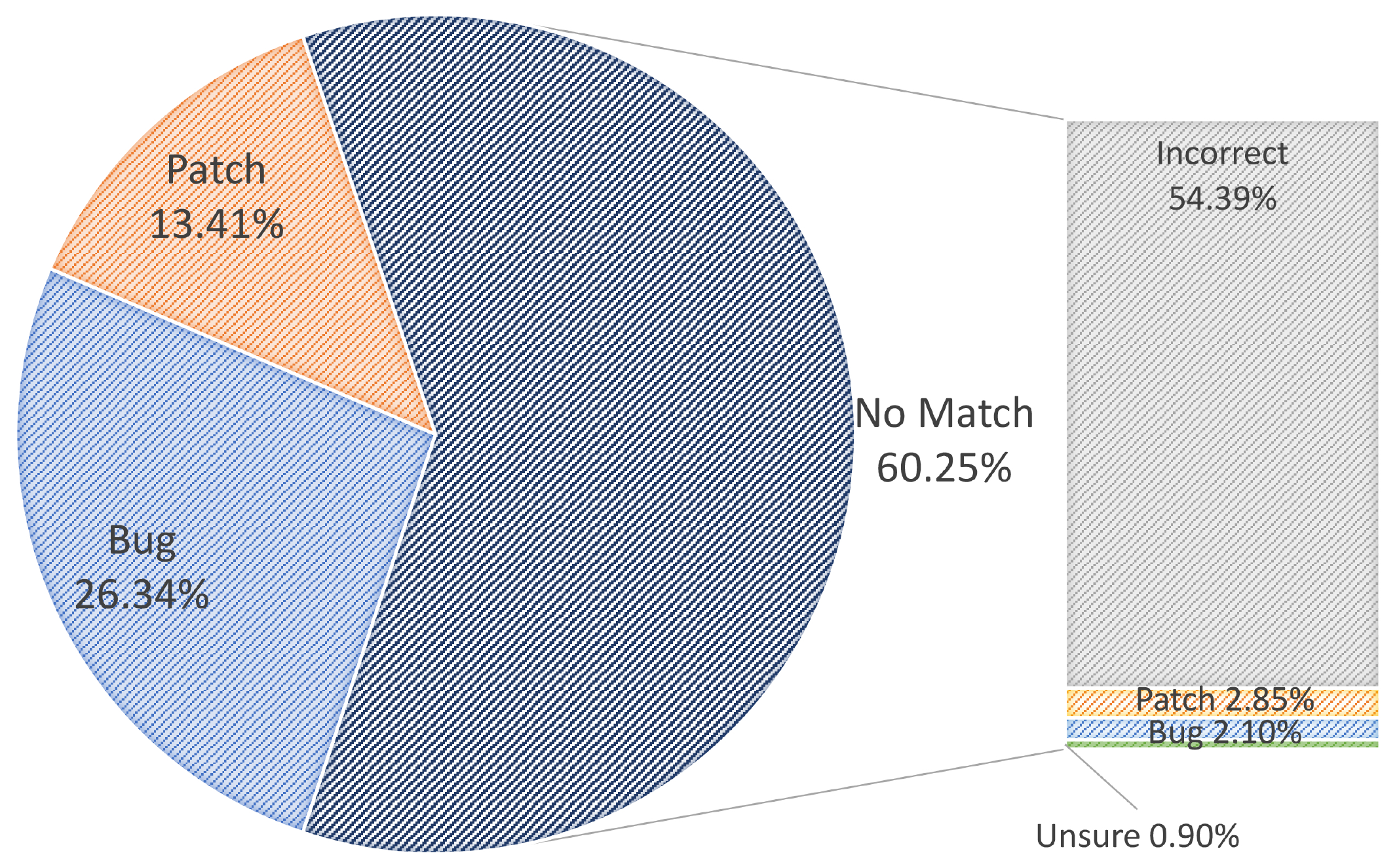}
  \caption{Match rate of Codex Davinci (left). Completions that do not match a patch or SStuB are validated by hand (right).}
  \label{fig:pie}
\end{figure}

\section{Methodology}
In this section, we describe the methodology for evaluating Codex\cite{chen2021evaluating}, PolyCoder\cite{xu2022systematic}, and CodeGen \cite{nijkamp2022conversational} on SStuBs. 

\subsection{Experimental Setup}

\noindent
\textit{ManySStuBs4J}

This dataset consists of a small and large dataset with 10,231 and 63,923 single-statement bug-fix changes (\emph{a.k.a} SStuBs) mined from 100 and 1000 popular repositories respectively. These SStuBs must match one of 16 bug templates. 
The goal is to collect bugs that are difficult to locate, but easy to fix. 
It's natural to wonder if automated coding tools based on language models could introduce such single-statement bugs.

For our study, we use the ManySStuBs4J \emph{large} dataset of 63,923 samples and use \texttt{\small git checkout} to obtain versions of the bug prior and after being fixed. We use \texttt{\small git blame} to determine when the bug was introduced and capture key statistics such as the number of commits to fix the SStuB. The bug locations within the files are indexed with fields \texttt{\small bugNodeStartChar} and \texttt{\small bugNodeLength}. Using these fields we can find the code before the bug, the bug, the fix, and the code after the bug/fix. Our experiment focuses on giving Codex a piece of code prior to the bug and seeing if Codex generates the correct code (fix) or incorrect code (bug). A large concern in this experiment is to make sure Codex has an equal opportunity to generate the known bug or patch. Therefore, we remove SStuBs that have other changes besides fixing the SStuB, which could otherwise \textit{condition} or bias Codex to make a decision; for example, a new variable only exists in the buggy version so Codex completes the bug. 
This leaves 34,595 bugs prior to deduplication. Then we drop duplicates for bugs that share the exact same prefix, bug, and fix. It is important to not inflate results by duplicate code examples\cite{allamanis2019adverse}. The remaining 16,899 SStuBs are used for evaluating all models.

\hfill \break
\noindent
\textit{Codex, PolyCoder, CodeGen}

Large language models like Codex, PolyCoder, and CodeGen  are demonstrably useful in code completion tools like Copilot, and are available for experimentation. Other models exist, like AI21 Jurassic; however they are not free, and would be costly at our scales, so were excluded from our study. We also didn't have the computational resources to run very large models locally. 
For these reasons, we follow the methodology from Xu \etal \cite{xu2022systematic} and use CodeGen, PolyCoder, and Codex.

Codex derives from GPT-3;  its training data consists of natural language and source code from available sources like public GitHub repositories. Codex has two sizes one called \textit{cushman-codex} and \textit{davinci-codex}\cite{openai_api}. To query  \textit{cushman-codex} and \textit{davinci-codex} models, we must make API requests using the OpenAI API (free, but rate-limited to 20 requests a minute). While the exact number of parameters is unknown, for both \textit{cushman-codex} and \textit{davinci-codex} models, prior work suggests sizes of 12B and 175B parameters\cite{rajkumar2022evaluating,chen2021evaluating,prenner2021automatic} for \textit{cushman-codex} and \textit{davinci-codex} respectively. Codex is primarily trained on Python\cite{chen2021evaluating}.

To determine if our results generalize to other large language models (LLMs) for code, we evaluate two additional families of models on SStuBs. These models are trained with different procedures and are readily available.  CodeGen\cite{nijkamp2022conversational} is an auto-regressive transformer model, trained with the next-token prediction objective on a corpus of code and natural language from GitHub. CodeGen is trained on multiple languages, but Python is the primary language.
PolyCoder\cite{xu2022systematic} is based on GPT-2 architecture and is trained on 250GB of code across 12 programming languages with C, C++, and Java being the primary language. PolyCoder outperforms all other code LLMs in C including Codex. Codex, PolyCoder, and CodeGen represent a diverse set of models all with several model size versions. Testing our the SStuBs hypothesis on Codex, PolyCoder, and CodeGen highlights potential risks of inducing SStuBs while using LLMs. While we cannot say for certain, other LLMs trained on similar data could show similar behavior.

We use the aforementioned models to generate completions by 
prompting with the code before the SStuB. When models complete the prompt, we can analyze the completion, by matching the known bug, or fix, from the SStuBs dataset. To compare completions to the bug and patch ground truths, we use substring matching (ignoring whitespace and formatting). To verify the accuracy of the results, a survey was conducted where the authors determined if sampled completions (n=401) match the bug, fix, or no match in a manner the automatic evaluation could not capture. For each model family, the best performing model completions were subjected to finer scrutiny; it is important that semantic equivalents are properly counted, such as Codex replacing a constant for the equivalent literal value. The manual survey interface\footnote{The annotation tool is a fork from localturk, a tool designed to emulate Amazon's Mechanical Turk. \url{https://github.com/danvk/localturk}} is screen-shot in \autoref{fig:survey}. \autoref{fig:survey:incorrect} shows a completion that is logically incorrect. In \autoref{fig:survey:correct}, Codex actually replaces a constant with its literal value which matches the fix. 401 randomly selected SStuBs are evaluated across each of the three model families to guarantee a confidence level of > 95\%. The overwhelming majority of completions are semantically incorrect to the bug or patch, see \autoref{fig:pie}.

\begin{lstlisting}[language=Java, caption={Prompting Codex with hint.}, label={prompt1}]
// Fix bugs in the below function.
...
g2d.setColor(tabFillColor);
g2d.fill(shaper.reset().doRect(boundsX, topY + shape.path.deltaY(1), 
    boundsWidth, paintBorder.top).getShape());
    
if (
\end{lstlisting}

\begin{lstlisting}[language=Java, caption={Prompting Codex the bug and fix.},label={prompt2}]
// Fix bugs in the below function

// Buggy Java
paintBorder.top >= 1

//Fixed Java
paintBorder.top > 1
...
g2d.setColor(tabFillColor);
g2d.fill(shaper.reset().doRect(boundsX, topY + shape.path.deltaY(1), 
    boundsWidth, paintBorder.top).getShape());
    
if (
\end{lstlisting}

\begin{figure}
    \centering
\includegraphics[width=1\linewidth]{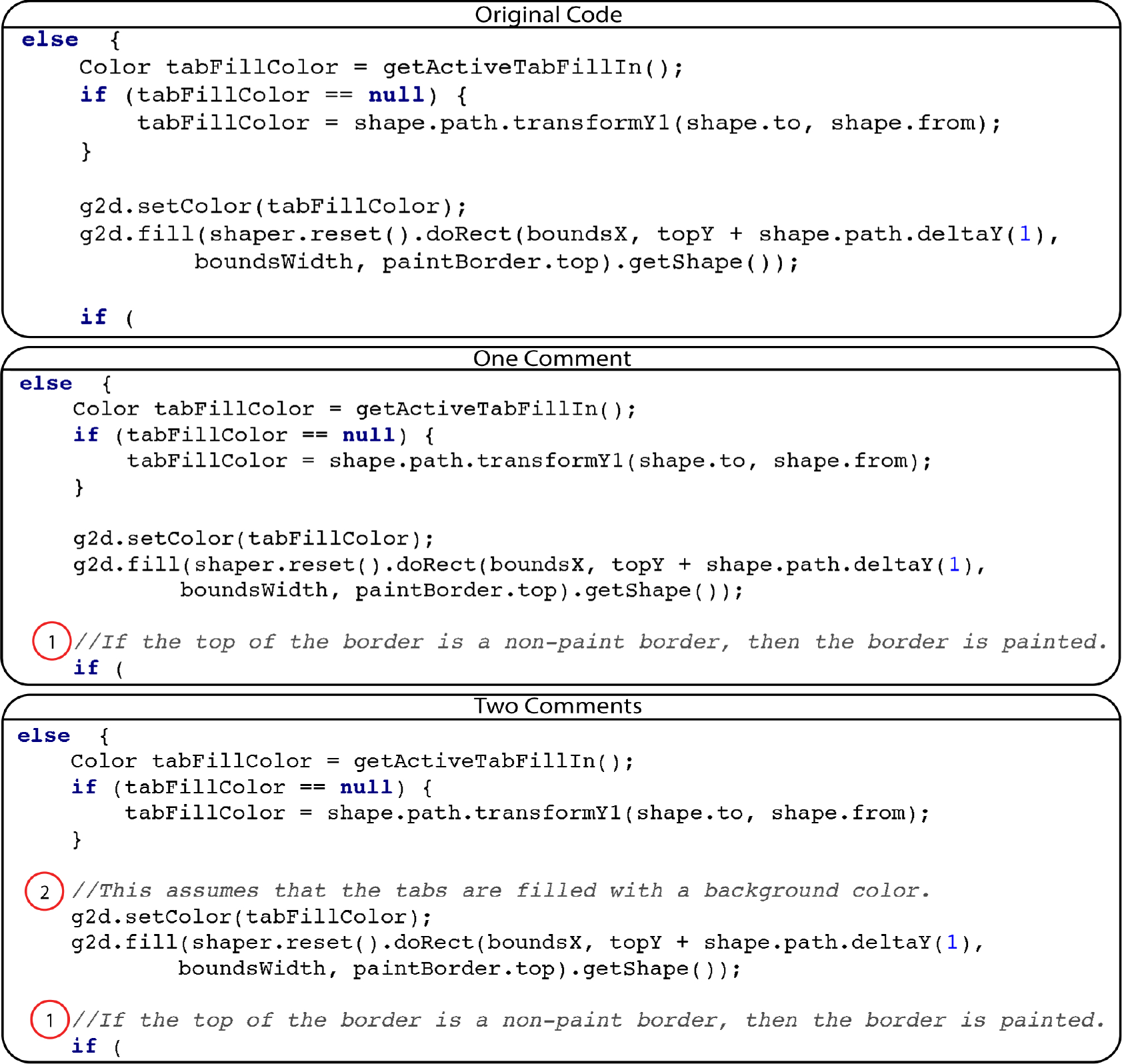}
  \caption{Adding neural-generated comments, step by step, in the prompt preceeding the  SStuB. The first added comment induces greatest improvement in generated code.}
  \label{fig:addingcomments}
\end{figure}
\hfill \break
\noindent
\textit{Prompting LLMs with Comments}

 Large language models were found to be surprisingly effective with good prompting\cite{liu2021pre}. 
``Prompt engineering'' is the process of constructing a text prompt, either just a textual prefix and/or set of explicit instructions to induce generation of desired text.
Prompt engineering has shown an effect on fixing programs, and generating solutions to coding questions\cite{wang2022no, pearce2021can, karmakar2022codex}; see \autoref{prompt1} for an example. Effective prompts may include sample input-output pairs\cite{prenner2021automatic} or SQL queries\cite{trummer2022codexdb,rajkumar2022evaluating}; \autoref{prompt2} is an example of bug-fixing comments according to the OpenAI API instructions\cite{openai_api}. This form of traditional hard-prompting\cite{wang2022no}, named hard because it uses hard-coded language, requires a \textit{prior} \emph{viz.} some known information about the input, \eg, it is buggy. In Codex experience surveys\cite{vaithilingam2022expectation}, it appears that prompt engineering is not useful \emph{in situ}, for actual coding. Still, prior work suggests prompt engineering can sometimes be useful~\cite{petroni2020context,prenner2021automatic, wang2022no}.

We hypothesize including comments within a code prompt
\emph{might} pre-condition the model better, to produce  more relevant and better overall code, in a couple ways:  
(1) generated code, if relevant, will be easier to understand; previous works show that comments help code comprehension~\cite{vaithilingam2022expectation, tenny1988program,woodfield1981effect}. (2) comments will help generated code be more maintainable~\cite{hartzman1993maintenance,lientz1983issues}. This led us to use comments to augment the readability and maintainability of the code prior to the SStuB for Codex. \autoref{fig:addingcomments} illustrates the incremental addition of comments starting around the SStuB, and then to surrounding code. We \emph{automatically} generate comments using CodeTrans comment generation model\cite{elnaggar2021codetrans}, trained on the DeepCom dataset\cite{hu2018deep}. The comment generation model\cite{elnaggar2021codetrans} can use any number of statements to condition its outputs on; we chose to use two statements for each comment, plus the buggy or fixed line, to keep the comment related to the SStuB.  

Comments can be generated from either the buggy or fixed version of code. Comments generated from the fixed version of the code should represent the correct logical steps through the single statement bug. On the other hand, comments generated from the buggy version should represent mostly correct steps with a minor single statement mistake. We test all models using \emph{both versions} of generated comments; the fixed version representing a ``non-buggy'' comment and the buggy version a ``buggy'' comment; this facilitates an evaluation of how non-buggy vs. buggy comments influence Codex's ability to avoid/make a single-statement mistake. We suspect that commenting in general, regardless of minor mistakes in natural language descriptions, will still condition Codex to use more reliable, well commented data for generation. \autoref{fig:addingcomments} shows the incremental addition of comments with automatic comment generation tools. We find that the comments \textit{are} beneficial for Codex models, irrespective of the developer's minor misunderstanding as conveyed in comments.

\hfill \break
\noindent
\textit{Prompt input, length, and SStuB completion}

The code prior to the SStuB, if available, is the conditional input to the model or prompt. The prompt, or code prior to the SStuB, is identical for all 16,899 SStuBs which guarantees the model is not biased towards the bug or the fix. When commenting the SStuB, an automatically generated comment from CodeTrans is placed above the lines used to generate it, properly tabulated, such that the comment appears natural as a developer would place it; see \autoref{fig:snippet1comment}. The prompt includes the code prior to the bug and up to the maximum allowed amount for each model. The length of any comments reduces the available input we can pass to the LLMs due to the fixed-length token window. The token window for Davinci is 8000 and the other models, Cushman, PolyCoder and CodeGen, are 2048. The token window is ultimately reduced further by the code completion length as the model performs generation in an auto-regressive fashion.

LLM code completions can span several lines. Codex \textit{can} use a stop token, such as the newline character, to terminate completion early. We found that LLMs, like Codex, often add arbitrary newlines and whitespace to completions; thus terminating completion on a newline might otherwise leave  unmatched completions. Instead, we ask the LLMs to complete a length of 64 tokens, which is sufficient for almost all SStuBs; SStuBs have a mean length 29 tokens and a median length 25 tokens. After generating a sequence of length 64, the completion is compared to the SStuB ignoring whitespace. The generated sequence \textit{must} match the SStuB completely to count as a bug or fix.


In the next section, we present the results from our findings: how often Codex and LLMs produce SStuBs, the number of commits to fix the generated SStuBs, and how annotating code with comments can improve performance on SStuBs.

\begin{table}
    \caption{SStuB production rate on off-the-shelf LLMs}
    \label{tabs:sstubrate}
    \begin{subtable}{\linewidth}
      \centering
        \caption{LLM completed SStuBs vs. correct code.}
                \resizebox{\textwidth}{!}{
    \begin{tabular}{cccccc}
\toprule
          Model &  Bugs & Patches & No Match &  Bug/Patch & Match Rate (\%)\\
\midrule
PolyCoder 160M  & 3429  &   1635  &     11835  &             2.10 &            29.97 \\
PolyCoder 0.4B  & 3672  &   1852  &     11375  &             1.98 &            32.69 \\
 PolyCoder 2.6B  & 3924  &   2096  &     10879  &             1.87 &            35.62 \\
    CodeGen 350  & 3709  &   1911  &     11279  &             1.94 &            33.26 \\
     CodeGen 2B  & 4102  &   2756  &     10041  &             1.49 &            40.58 \\
     CodeGen 6B  & 4168  &   2944  &      9787  &             1.42 &            42.09 \\
    CodeGen 16B  & 4299  &   3296  &      9304  &             1.30 &            44.94 \\
   Cushman 12B  & 3775  &   1833  &     11291  &             2.06 &            33.19 \\
  Davinci 175B  & 4452  &   2267  &     10180  &             1.96 &            39.76 \\
\bottomrule
\end{tabular}
    }
    \label{tab:sstubrate}
    \end{subtable}%
    \\\\\\
    \begin{subtable}{\linewidth}
      \centering
        \caption{Manually examined model predictions when neither bug or patch \textit{a.k.a.} ``No Match'' is detected.}
\npdecimalsign{.}
\nprounddigits{2}
    \resizebox{\textwidth}{!}{
\begin{tabular}{cc|c n{2}{2} |c n{2}{2} |c n{2}{2} }
\toprule
\multirow{1}{*}{\bfseries Model} &&
    \multicolumn{2}{c}{\bfseries PolyCoder 2B}&
    \multicolumn{2}{c}{\bfseries CodeGen 16B }&
\multicolumn{2}{c}{\bfseries Davinci}\\

    & &  counts &        \% &  counts &        \% &  counts &        \% \\
\midrule
Incorrect &&     361 & 90.024938 &     357 & 89.027431 &     362 & 90.274314 \\
      Patch & &     19 &  4.738155 &      28 &  6.982544 &      19 &  4.738155 \\
    Bug &&      15 &  3.740648 &      10 &  2.493766 &      14 &  3.491272 \\
   Unsure & &      6 &  1.496259 &       6 &  1.496259 &       6 &  1.496259 \\
\bottomrule
\end{tabular}
 }
    \label{tab:sstubsurvey}
    \end{subtable} 
    
\end{table}

\begin{figure}
  \centering
  \includegraphics[width=.8\linewidth]{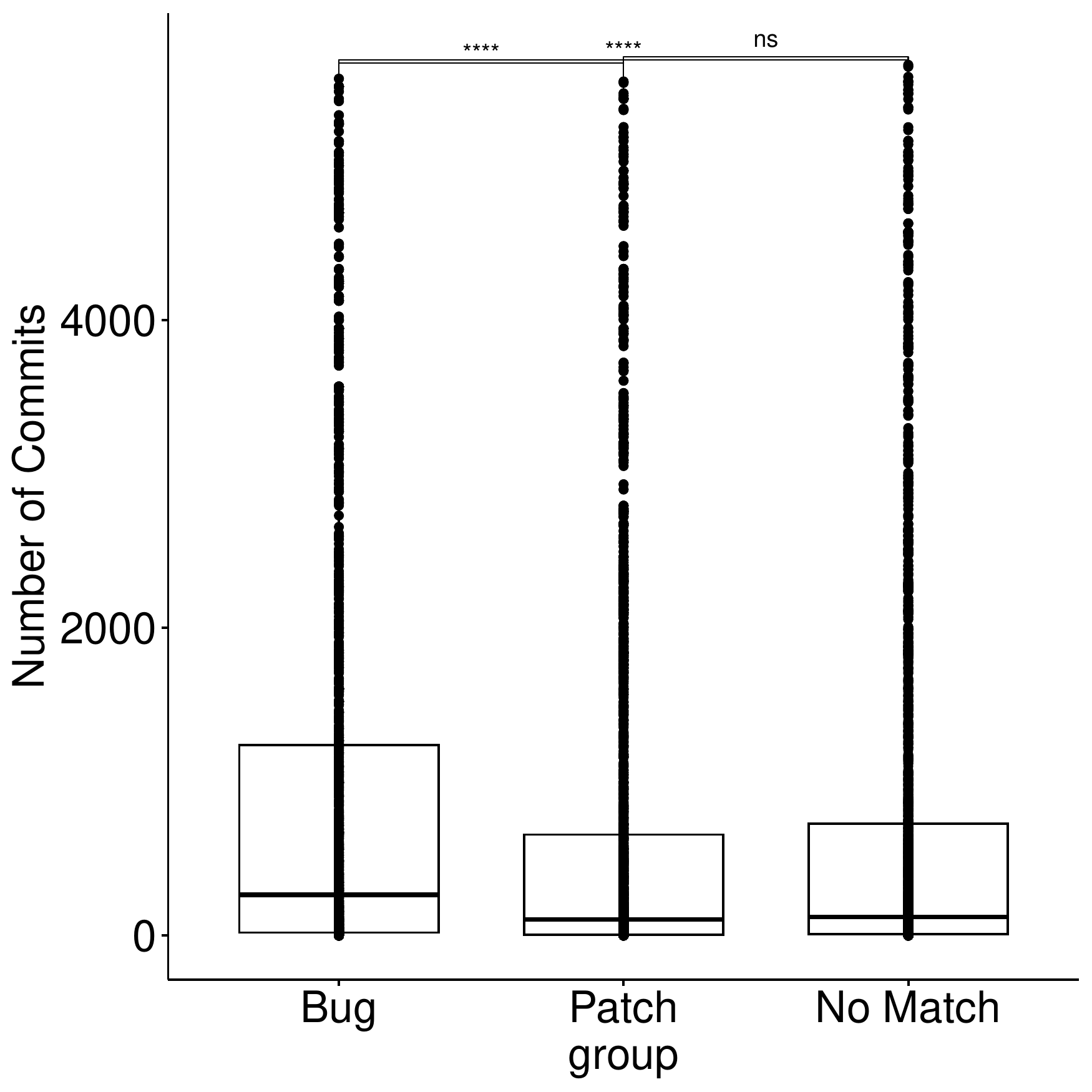}
  \caption{Developers take more time, measured in commits, to resolve SStuBs that Codex generates. All differences are pairwise statistically significant to p $\leq$ 0.0001.}
  \label{fig:committime}
\end{figure}

\section{Results}
\subsection{SStuB production in LLMs (RQ1)}
\autoref{tabs:sstubrate} is the number of bugs, patches, and non-matching completions from studied LLMs. We use the bug/patch ratio, as a metric to universally compare models as the overall number of successful completions, either a SStuB or a patch, varies between models. Codex, PolyCoder, and CodeGen 350M all produce nearly 2x as many bugs as patches. \textit{Davinci-codex} and \textit{cushman-codex} perform surprisingly poorly given their size, and extensive training. It is plausible that the Codex models recapitulate bugs seen in training data\cite{karmakar2022codex}, however, many of these SStuBs will have been addressed per their inclusion in the ManySStuBs4J corpus two years ago; \emph{viz.} there is a fix. Although Codex produces a high rate of SStuBs, Codex is capable of avoiding 13.41\% of SStuBs in the dataset.

\vspace{0.1in}

\noindent
\setlength{\fboxsep}{2pt}
\setlength{\fboxrule}{1pt}
\fcolorbox{black!60}{gray!20}{%
    \parbox{\columnwidth}{
 \textbf{RQ1}: Codex and LLMs produce twice as many SStuBs as correct code. Codex manages to avoid 13.41\% of SStuBs.
 }
    }

\begin{table*}
\centering
\caption{Bug Rate and Patch Rate after adding a comment around the SSTUB.}
\label{tab:commentdiff}
\resizebox{\textwidth}{!}{
\begin{tabular}{|lc|cc|cc|cc|c|}
\toprule
Model Name
& \multicolumn{1}{p{2cm}}{\centering Model Size \\(Billions)}
& \multicolumn{1}{|p{2cm}}{\centering Bug Change}

& \multicolumn{1}{p{2cm}|}{\centering \% Change}
& \multicolumn{1}{p{2cm}}{\centering Patch Change}

& \multicolumn{1}{p{2cm}}{\centering \% Change}
& \multicolumn{1}{|p{2cm}}{\centering Bug/Patch \\ Ratio Change}
& \multicolumn{1}{p{2cm}|}{\centering \% Change}
& \multicolumn{1}{p{2cm}|}{\centering Match Rate\\Change}\\
\midrule
PolyCoder 160M &       0.16 &  -427 &    -12.45 &      150 &      9.17 &                   -0.42 &    -20.42 &            -1.64 \\
PolyCoder 400M &        0.4 &  -376 &    -10.24 &      250 &     13.50 &                   -0.41 &    -21.94 &            -0.75 \\
PolyCoder 2.6B &        2.6 &  -407 &    -10.37 &      323 &     15.41 &                   -0.42 &    -23.23 &            -0.50 \\
  CodeGen 350M &       0.35 &  -427 &    -11.51 &      256 &     13.40 &                   -0.43 &    -21.70 &            -1.01 \\
   CodeGen 2.7 &        2.7 &  -655 &    -15.97 &      129 &      4.68 &                   -0.29 &    -18.73 &            -3.11 \\
    CodeGen 6B &        6.1 &  -742 &    -17.80 &     -174 &     -5.91 &                   -0.18 &    -11.59 &            -5.42 \\
   CodeGen 16B &       16.1 &  -759 &    -17.66 &      -81 &     -2.46 &                   -0.20 &    -10.24 &            -4.97 \\
 Codex Cushman &       12.0 &   800 &     21.19 &     2329 &    127.06 &                   -0.96 &    -44.90 &            18.52 \\
 Codex Davinci &      175.0 &   877 &     19.70 &     2882 &    127.13 &                   -0.93 &    -46.08 &            22.24 \\
 \bottomrule
\end{tabular}
}
\end{table*}

\begin{table}[]
    \caption{LLM completed SStuBs vs correct code with a comment prior.}
    \label{tab:sstubratecomments}
    \centering
        \resizebox{.5\textwidth}{!}{
    \begin{tabular}{cccccc}
\toprule
          Model &  Bugs & Patches & No Match &  Bug\textbackslash Patch & Match Rate (\%)\\

\midrule
PolyCoder 160M  &  3002 &     1785 &       12112 &             1.68 &            28.33 \\
PolyCoder 0.4B  &  3296 &     2102 &       11501 &             1.57 &            31.94 \\
 PolyCoder 2.6B  &  3517 &     2419 &       10963 &             1.45 &            35.13 \\
    CodeGen 350M  &  3282 &     2167 &       11450 &             1.51 &            32.24 \\
     CodeGen 2B  &  3447 &     2885 &       10567 &             1.19 &            37.47 \\
     CodeGen 6B  &  3426 &     2770 &       10703 &             1.24 &            36.66 \\
    CodeGen 16B  &  3540 &     3215 &       10144 &             1.10 &            39.97 \\
   Cushman 12B  &  4575 &     4162 &        8162 &             1.10 &            51.70 \\
  Davinci 175B  &  5329 &     5149 &        6421 &             1.03 &            62.00 \\
\bottomrule
\end{tabular}
}
\end{table}

\subsection{Number of commits to fix LLM produced SStuBs (RQ2)}
\autoref{fig:committime} shows the number of commits to fix of SStuBs where Codex generates the original human-created `Bug', or a `Patch', or something else (`No Match'). For each of these categories, we examine the version-control history to examine how long (count of commits from introduction
to fix, using {\small\tt git blame}) developers took to fix them. 
Unfortunately, the number of commits to fix when Codex (re)produces SStuBs (bugs) \emph{is significantly longer} than in other cases. The median number of commits to fix for the bugs, patches, and no match is 265, 106, and 121 commits respectively. Significance of pairwise t-tests was sustained even after the conservative Bonferroni correction. This finding suggests that \emph{when Codex generates  SStuBs, these might inherently take human developers longer to fix}! If used widely in open-source code, Codex might spout SStuBs that live longer (in version history) and further pollute future Codex training data. We believe future, detailed investigation in the 4452 matching SStuBs might help improve Codex. 
\\
\\
\noindent
\setlength{\fboxsep}{2pt}
\setlength{\fboxrule}{1pt}
\fcolorbox{black!60}{gray!20}{%
    \parbox{\columnwidth}{
 \textbf{RQ2}: The ManySStuBs4J data suggest that in cases where Codex wrongly generates simple, stupid bugs, these may take developers significantly longer to fix than in cases where Codex doesn't. 
 }
    }

\begin{figure}

  \centering
  \includegraphics[width=.8\linewidth]{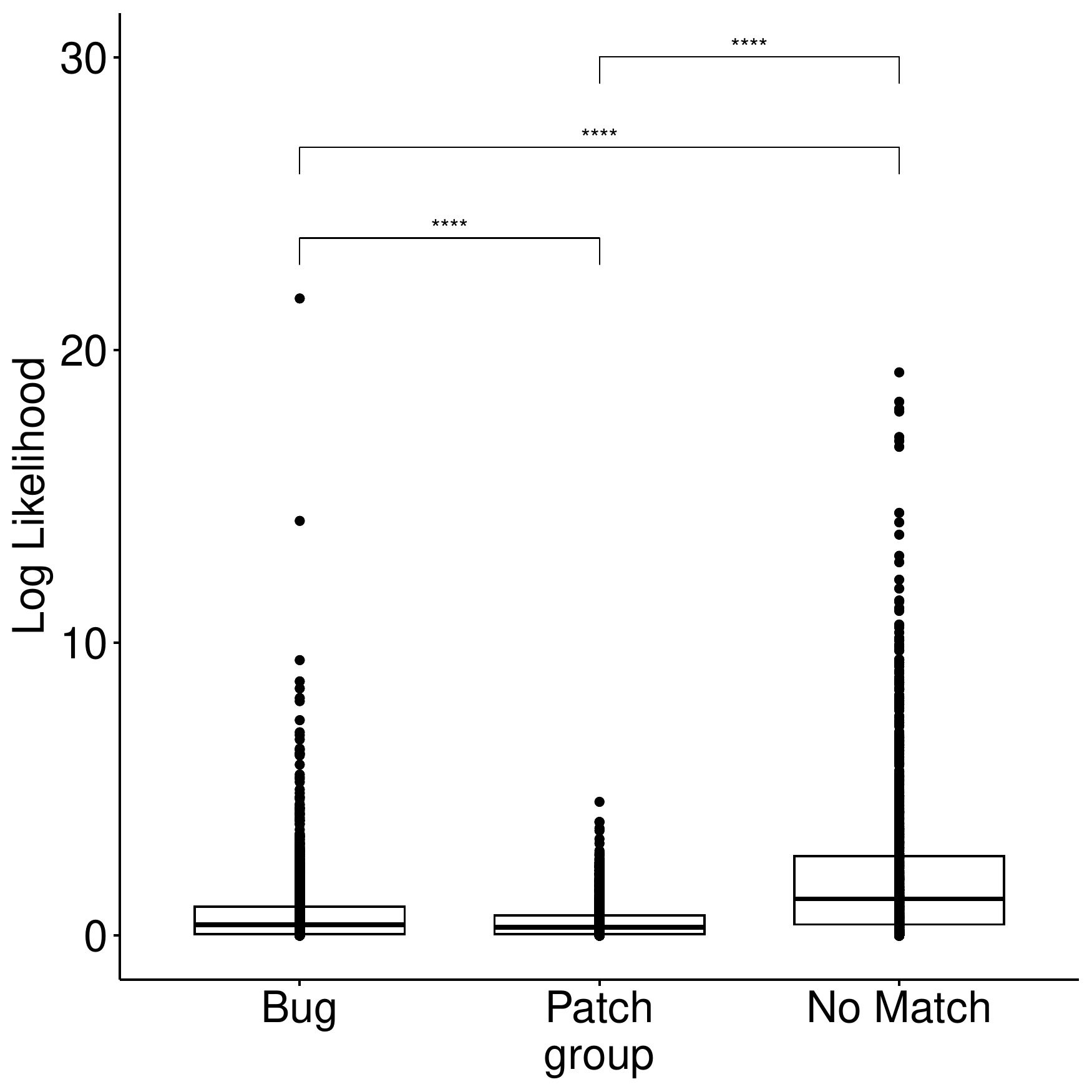}
  \caption{SStuB bugs and patches from Codex are equally more\\ natural than other code it generates. All differences are pairwise statistically significant to p $\leq$ 0.0001.}
  \label{fig:likelihood}
\end{figure}%

\subsection{SStuB regularity (RQ3)}
The significant number of commits to fix SStuBs vs. patches (RQ2) motivates a comparison of the ``naturalness'' of bugs, patches, and no match group of SStuBs. \autoref{fig:likelihood} shows that there is little difference between the negative log-likelihood of bugs and patches. As expected, the `no-matches' have a higher negative log-likelihood, since  these completions were presumably not seen in the training set. 
The similar negative log-likelihood of SSTuBs and patches suggests
that it may be challenging to fine-tune Codex to detect or
avoid SStuBs, since Codex rates them both equally `natural';
we leave this for future work. However we do study if proper prompt engineering (\emph{e.g.,} with comments), might help matters. 
\\
\\
\noindent
\setlength{\fboxsep}{2pt}
\setlength{\fboxrule}{1pt}
\fcolorbox{black!60}{gray!20}{%
    \parbox{\columnwidth}{
 \textbf{RQ3}: Codex log-probabilities indicate that SStuBs (bugs and patches) are regular and natural, thus making detection difficult.
 }
    }

\begin{figure}
\centering
\begin{subfigure}{.5\textwidth}
  \centering
  \includegraphics[width=\linewidth]{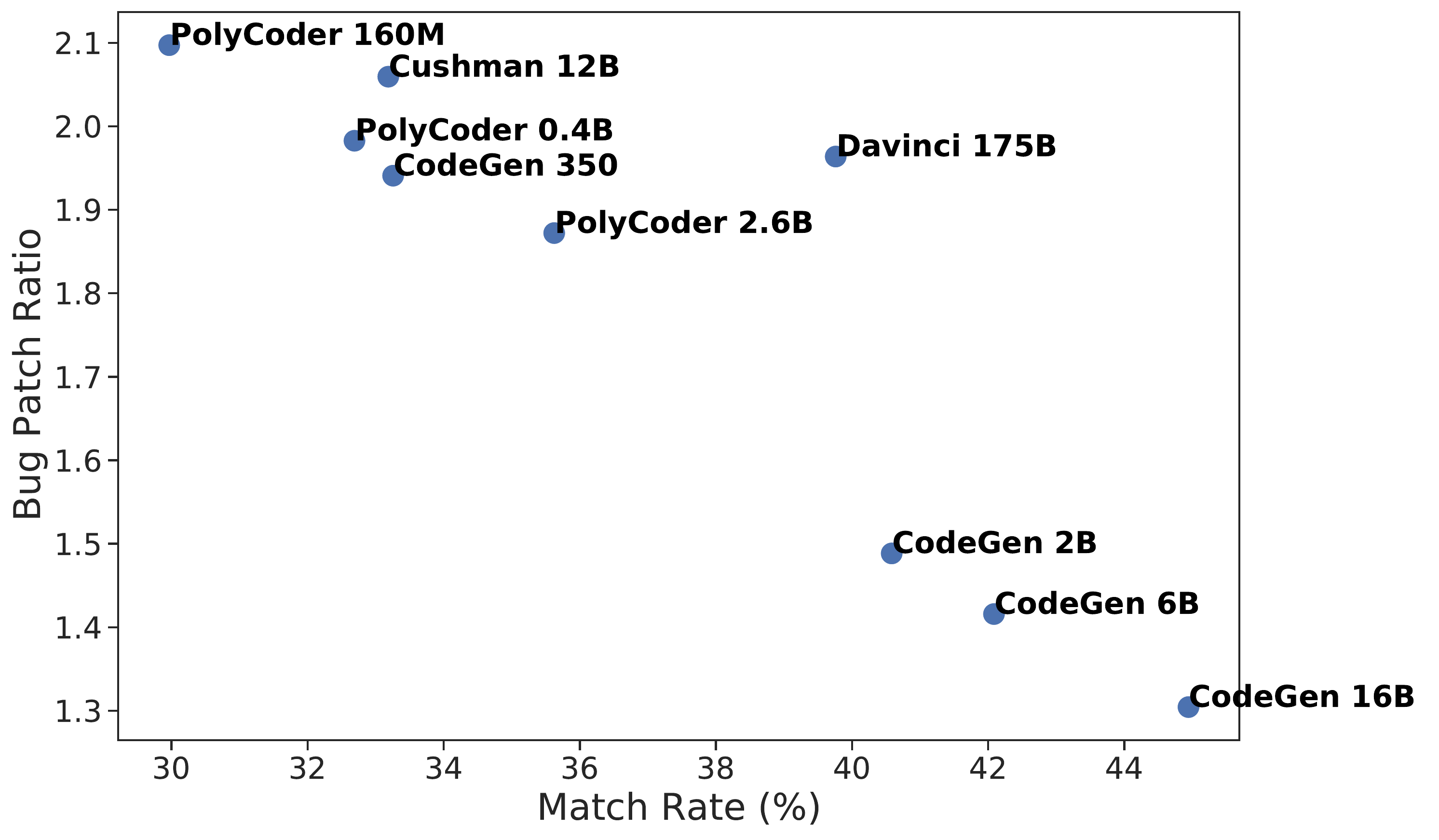}
  \caption{Codex models (Cushman \& Davinci), without comments, perform not well on bug/patch ratio and the match rate.}
  \label{fig:beforescatter}
\end{subfigure}%
\vspace{1mm}
\begin{subfigure}{.5\textwidth}
  \centering
  \includegraphics[width=\linewidth]{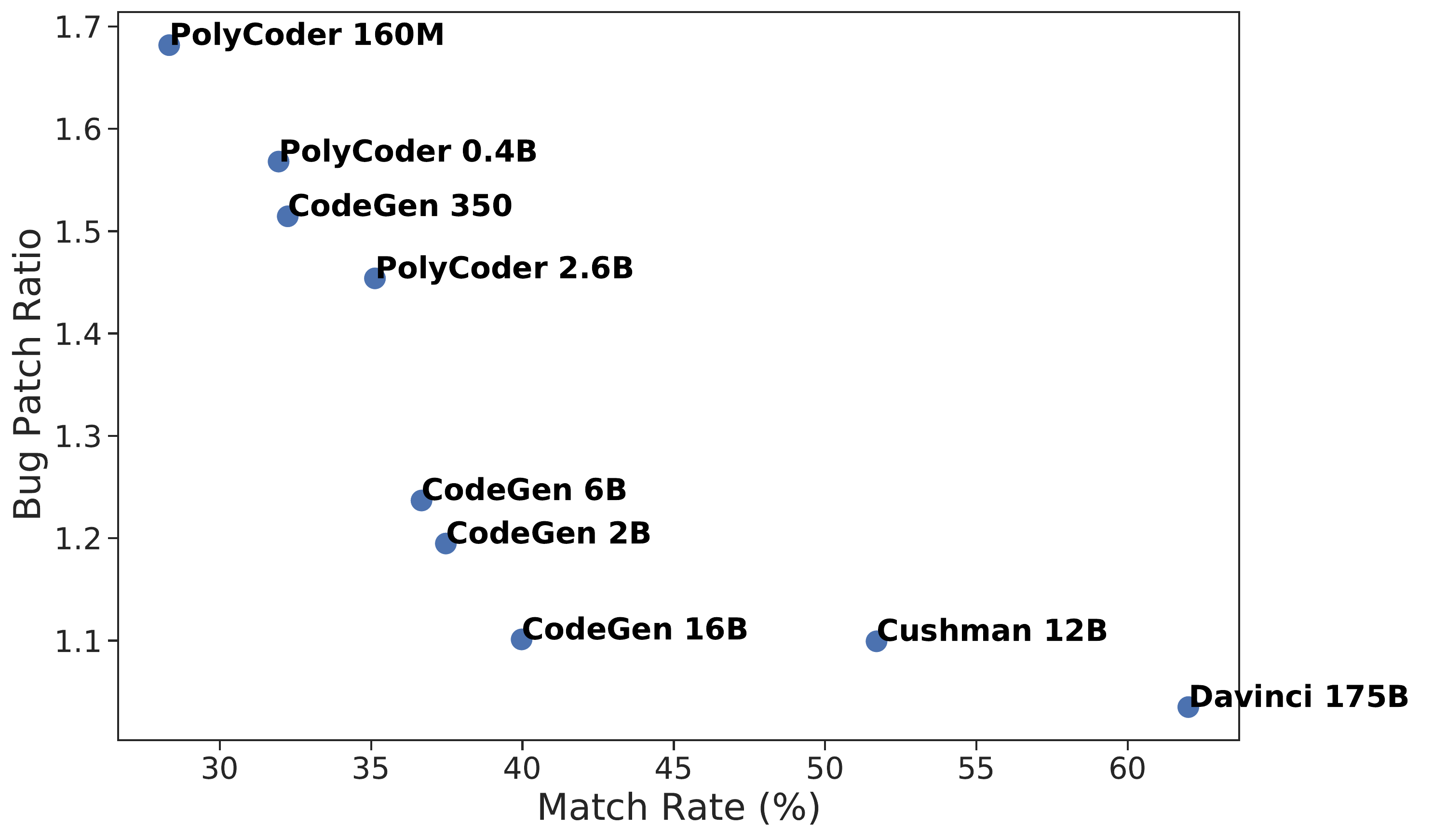}
  \caption{Commenting code helps Codex achive best performance across SStuBs while improving the match rate; }
  \label{fig:afterscatter}
\end{subfigure}
\caption{Prompting with comments should be used to both avoid SStuBs}
\label{fig:scatter}
\end{figure}

\begin{figure}
\centering

\includegraphics[width=\linewidth]{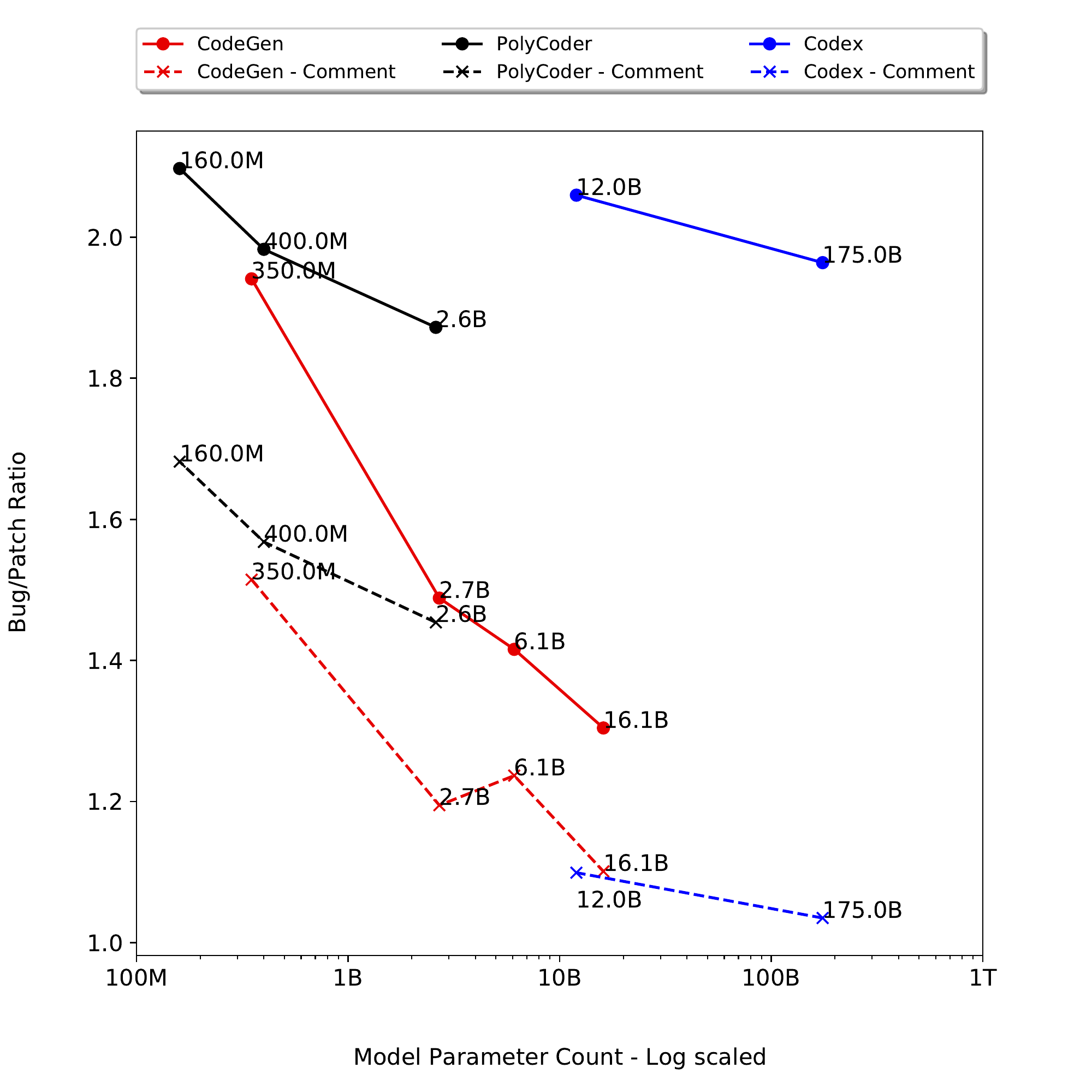}
  \caption{Bug/Patch Ratio vs. Parameter Count. Three model families at various sizes. The largest difference is the addition of 1 comment prior to the SStuB.}
  \label{fig:paramvs}

\end{figure}

\subsection{Avoiding SStuBs (RQ4)}
We now turn to the question of whether adding natural language comments to the prompt suppresses SStuB generation by Codex. \autoref{tab:sstubratecomments} shows the results of inserting a single comment into the prompt. \autoref{tab:commentdiff} captures the rate change in bugs, patches, and bug/patch ratio after adding a comment prior to the SStuB. First, Codex and other LLMs PolyCoder and CodeGen behave differently, namely in the match rate; Codex match rate increases by 19-22\% where as other LLMs do not change much. In PolyCoder and CodeGen, the number of bugs decreases from 10-18\% and the patch rate increases. The bug/patch ratio in all models improves! Codex Cushman and Davinci generate 20\% more bugs, but then produce \textit{127\%} more patches. The bug/patch ratio is cut by almost half. This suggests that developers get better results by commenting code while using Codex: this amounts to about 3000 more patches (5149 vs. 2267). 

\autoref{fig:scatter} are scatter plots showing the relationship between match rate and bug/patch ratio. Ideally, models will have a high match rate (less
unknown cases) and a low bug/patch ratio. Per \autoref{fig:beforescatter}, Codex models Cushman and Davinci were not competitive to other off the shelf models. After adding comments to the SStuB prone code, \autoref{fig:afterscatter}, all models perform better and Codex performs \emph{much} better than the next best model CodeGen 16B.


\begin{table*}
\centering

\caption{Bug Rate and Patch Rate after adding erroneous comments around SSTUB.}
\resizebox{\textwidth}{!}{
\begin{tabular}{|lc|cc|cc|cc|c|}
\toprule
Model Name
& \multicolumn{1}{p{2cm}}{\centering Model Size \\(Billions)}
& \multicolumn{1}{|p{2cm}}{\centering Bug Change}

& \multicolumn{1}{p{2cm}|}{\centering \% Change}
& \multicolumn{1}{p{2cm}}{\centering Patch Change}

& \multicolumn{1}{p{2cm}}{\centering \% Change}
& \multicolumn{1}{|p{2cm}}{\centering Bug/Patch \\ Ratio Change}
& \multicolumn{1}{p{2cm}|}{\centering \% Change}
& \multicolumn{1}{p{2cm}|}{\centering Match Rate\\Change}\\

\midrule
PolyCoder 160M &       0.16 &    59 &      1.72 &     -237 &    -14.50 &                    0.40 &     18.97 &            -1.05 \\
PolyCoder 400M &        0.4 &   171 &      4.66 &     -132 &     -7.13 &                    0.25 &     12.69 &             0.23 \\
PolyCoder 2.6B &        2.6 &   234 &      5.96 &     -145 &     -6.92 &                    0.26 &     13.84 &             0.53 \\
  CodeGen 350M &       0.35 &   110 &      2.97 &     -178 &     -9.31 &                    0.26 &     13.54 &            -0.40 \\
   CodeGen 2.7 &        2.7 &   -78 &     -1.90 &     -420 &    -15.24 &                    0.23 &     15.74 &            -2.95 \\
    CodeGen 6B &        6.1 &  -264 &     -6.33 &     -603 &    -20.48 &                    0.25 &     17.79 &            -5.13 \\
   CodeGen 16B &       16.1 &   -95 &     -2.21 &     -558 &    -16.93 &                    0.23 &     17.72 &            -3.86 \\
 Codex Cushman &       12.0 &  1511 &     40.03 &     1135 &     61.92 &                   -0.28 &    -13.52 &            15.66 \\
 Codex Davinci &      175.0 &  1620 &     36.39 &     1613 &     71.15 &                   -0.40 &    -20.31 &            19.13 \\
\bottomrule
\end{tabular}
}
\label{tab:diffbadcomment}
\end{table*}

\begin{figure*}
\centering
\begin{subfigure}{.45\textwidth}
  \centering
  \includegraphics[width=.98\linewidth]{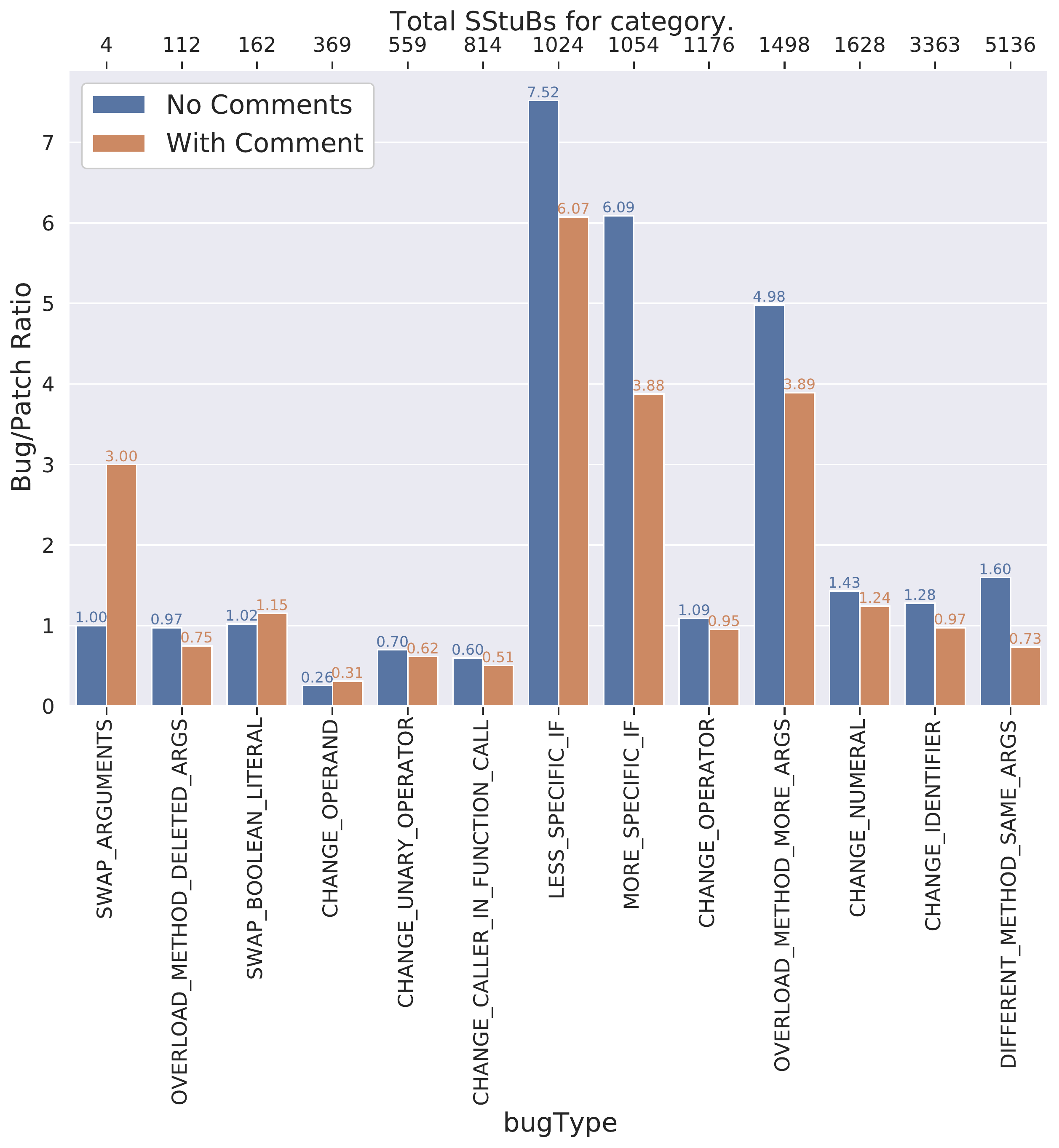}
  \caption{Comment effect on bug/patch ratio. Lower is better. Top axis is total SStuB count, little significance placed on bug categories with less than 100 samples. }
  \label{fig:bybugratio}
\end{subfigure}%
\hspace{1cm}
\begin{subfigure}{.45\textwidth}
  \centering
  \includegraphics[width=1.01\linewidth]{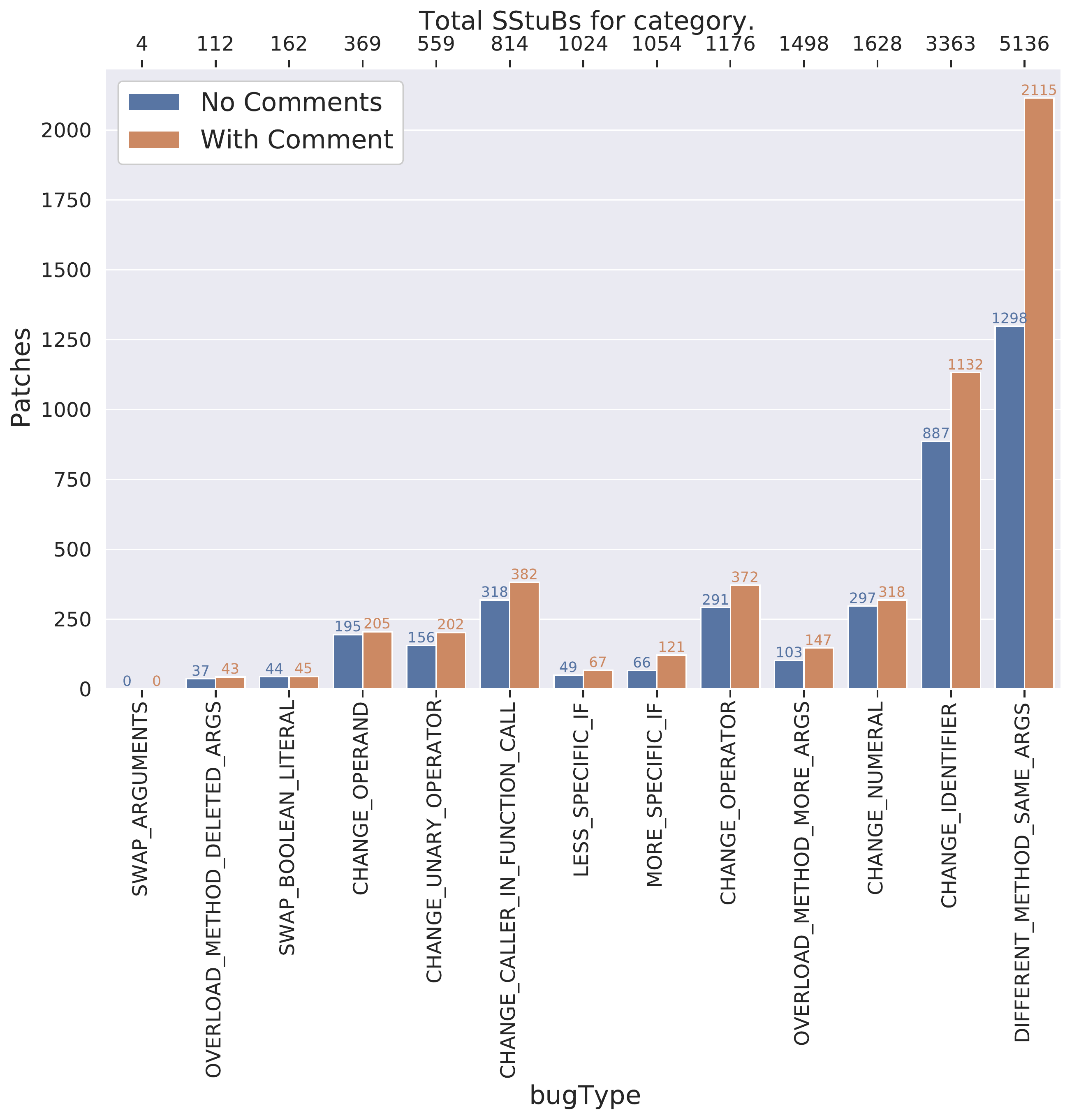}
  \caption{Comment effect on patches. Higher is better. Top axis is total SStuB count, little significance placed on bug categories with less than 100 samples. }
  \label{fig:bybugpatches}
\end{subfigure}
\caption{Effect of adding comments to Codex Davinci prompt.}
\label{fig:bybug}
\end{figure*}

\begin{table}[]
    \caption{Codex (Davinci) performance across various prompting techniques.}
    \label{tab:variouspromptingtechniques}
    \centering
    \resizebox{.5\textwidth}{!}{
    \begin{tabular}{cccccc}
\toprule
          Prompt &  Bugs & Patches & No Match &  Bug/Patch& Match Rate (\%)\\

\midrule
Hint  &  6228 & 4814 &   5857     &     1.29         & 65.34             \\
Bug \& Fix & 6565 & 6055 &    4279    &        1.08    &     74.68         \\
Comment &  5329 &     5149 &        6421 &             1.03 &             62 \\
\bottomrule
\end{tabular}
}
\end{table}

\autoref{fig:paramvs} shows the effect of adding comments to the bug/patch ratio with model parameter counts. Adding a comment in the prompt helps more than increasing parameter counts! A 160M parameter PolyCoder with a comment outperforms (by 14\% improvement in bug/patch ratio) both the Codex Cushman 12B and Davinci 175B, without comments. 

Finally, \autoref{fig:bybug} shows bug/patch ratio (\autoref{fig:bybugratio}) and number of patches (\autoref{fig:bybugpatches}) by the bug type (given in the ManySStuBs4J dataset) ranked left to right by the frequency in the dataset. The largest improvements in bug/patch ratio in a sufficient set of samples are \texttt{\small LESS\_SPECIFIC\_IF}, \texttt{\small MORE\_SPECIFIC\_IF}, \texttt{\small OVERLOAD\_METHOD\_MORE\_ARGS}, and \texttt{\small DIFFERENT\_METHOD\_SAME\_ARGS}. The bug types that gained the most patches are \texttt{\small CHANGE\_IDENTIFIER} and \texttt{\small DIFFERENT\_METHOD\_SAME\_ARGS}. \texttt{\small SWAP\_ARGUMENTS} got worse, but only consists of 4 total examples, and it is hard to make any conclusions from this. 

\subsection{Commenting vs Traditional Prompting (RQ4 cont.)}
Prompting LLMs both in NLP and SE applications often come in the form of instructions prior to the input. Previous empirical studies of Codex \cite{wang2022no,yetistiren2022assessing,sobania2022choose, prenner2021automatic,pearce2021can,pearce2022examining,pearce2022asleep,nguyen2022empirical,perry2022users,sandoval2022security} prompt Codex with instructions, input-output pairs, or examples prior to the input that the model is conditioning the output on. We compare in a similar fashion to Prenner \etal \cite{prenner2021automatic} by providing Codex with hints (\autoref{prompt1}), and SStuBs (\autoref{prompt2}). We remind the reader that both prompting techniques found commonly in previous works, require knowing something about the snippet of code that will be generated; \eg, a location in the code that has a SStuB (hint) and what the SStuB is and how to fix it (bug and fix).

While both approaches improve the bug/patch ratio and match rate, \autoref{tab:variouspromptingtechniques}, neither does as well as adding natural language comments. Furthermore the number of bugs greatly increases in the traditional prompting techniques.
\\
\\
\noindent
\setlength{\fboxsep}{2pt}
\setlength{\fboxrule}{1pt}
\fcolorbox{black!60}{gray!20}{%
    \parbox{\columnwidth}{
 \textbf{RQ4}: Commenting code can lead to less generated SStuBs and more generated patches. Codex models improve the most from code comments.
 }
    }

\subsection{What if we insert a `buggy' comment? (RQ5)}
Finally, we try inserting a \emph{buggy} comment in the prompt, to mimic the developers description of the SStuB. The natural language comment for buggy code is created by conditioning CodeTrans on the SStuB (bug) rather than correct code (patch). \autoref{tab:diffbadcomment} shows the difference in bugs, patches, bug/patch ratio, and match rate with a buggy comment before the SStuB. Surprisingly, Codex is robust and still improves bug/patch ratio over the ``no comments'' case. This suggests that the mere presence of relevant comments in the prompt sufficiently pushes the model to produce better code. It's also interesting to note that the lower capacity models (and also the 16B parameter CodeGen) tend to be misled by the `buggy' comments, whereas the larger capacity, well-trained Codex models are not.  
\\
\\
\noindent
\setlength{\fboxsep}{2pt}
\setlength{\fboxrule}{1pt}
\fcolorbox{black!60}{gray!20}{%
    \parbox{\columnwidth}{
 \textbf{RQ5}: Misleading comments still condition Codex to produce less SStuBs. Commenting appears beneficial irrespective of the developer's understanding of the SStuB.
 }
    }

\section{Discussion}

\subsection{Implications of Findings}
\noindent
\textit{Implications of Codex Producing SStuBs}

The good news: Our study suggests that LLMs like Codex do help avoid a significant number of SStuBs in our dataset, out-of-the-box,
and even more with non-buggy comments! 
But they \underline{do} produce simple, stupid bugs. 
Even Codex produces up to 2x more SStuBs as patches, if used directly, nor does increasing model size (see Table I) necessarily help. 

To better understand why Codex still produce SStuBs, one must further examine the training data (sadly, not available for many LLMs); we hope training data will become more available. Previous work \cite{pearce2022asleep,asare2022github, sobania2022choose} studying Codex-generated vulnerabilities blames Codex's language modeling roots, which push it to produce the most ``likely completion (for a given prompt) based on the encountered samples during training''. Also, SStuBs are capable of lasting for long periods of time\cite{mosolygo2021rise} and are not detected by continuous integration\cite{latendresse2021effective} or static analysis\cite{karampatsis2020often,mosolygo2021rise,hua2021effectiveness} which explains why Codex recapitulates them (from it's training data). 
Simple, stupid bugs are likely regularly injected by devs; training Codex without SStuBs would be challenging, given training data is drawn from 54 million repositories~\cite{chen2021evaluating}. The effect of Codex produced SStuBs is significant, and troubling. The number of commits to fix Codex produced SStuBs versus the avoided SStuBs is significant, taking more than twice as long to fix. Still we should bear in mind that Codex avoids 2,267 bugs on its own or 13.41\% of the dataset, indicating an AI paired programmer is helpful in avoiding SStuBs too.


\hfill \break
\noindent
\textit{Avoiding SStuBs with LLM}

Codex and other LLMs respond unpredictably to prompts, and developers often struggle to get LLMs to generate desired code~\cite{vaithilingam2022expectation}. Studies suggest that breaking coding tasks into manageable sub-problems helps~\cite{vaithilingam2022expectation, trummer2022codexdb}. NLP tasks work similarly; chain-of-thought\cite{wei2022chain} and reasoning step-by-step\cite{kojima2022large} improve problem-solving rate. Commenting is ideally a form of step-by-step reasoning, explaining high level steps, or clarifying confusing code. The generated comments we used appear to be high-level descriptions and not deep technical commentary of computed values and algorithmic mechanisms. Our work suggests minimal effort techniques, like automatically generated documentation, may help avoid SStuBs when using an AI programming assistant like Copilot. Not only can comments be automatically generated as documentation, but comments can be used directly as a prompt for Codex. To the best of our understanding, Codex might condition the generated code on a smaller search-space of non-buggy solutions, thus helping the developer avoid introducing SStuBs. 


Lastly, comments can sometimes be used to check the implementation consistency given the desired functionality\cite{wen2019large}. Future work could examine if SStuBs can be detected with the same tools given a set of generated comments. 

In our experiments the placement of comments is uniform, and further work should be done to determine best possible comment placement in a density that is adequate and not excessive; automatic methods exist \cite{huang2019learning}. Excessive commenting is typically symptomatic of a lack of understanding of the code and a ``code smell''\cite{becker1999refactoring}.



%



\hfill \break
\noindent
\textit{Maintaining AI Generated Code}

Language models for code like Codex, PolyCoder, CodeGen, and others\cite{feng2020codebert,guo2020graphcodebert,wang2021codet5,ahmad2021unified,kanade2020learning} will become bigger, and better at code completion~\cite{hellendoorn2021growing}. In a world where AI programming assistants learn from data at scale\cite{lachaux2020unsupervised}, it is hard to say how much of novel programming projects or code reuse\cite{allamanis2019adverse} will guide such tools. Fundamentally, there is a need for improved readability and comprehensibility in AI-generated code\cite{vaithilingam2022expectation}.  Code comments can improve comprehensibility of inserted code, especially of more difficult statements. Code that is more readable and understandable is much more maintainable. Our work suggests that comments help avoid SStuBs, in addition to the traditional role of improving code readability. Prior work indicates that SStuBs are usually \emph{not} fixed by the inserting developer, but by other developers\cite{zhu2021mea}, with greater effort; we note that a SStuB inserted by an AI programming assistant, is always an ``other developer''. 

 Lastly, the preliminary successes on SStuBs warrants further research in comment generation with AI programming assistants. Comment generation models like DeepCom\cite{hu2018deep} and CodeTrans\cite{elnaggar2021codetrans} are fully automated and could function in a variety of roles for AI programming assistants. For example, comment generation models could serve as automatic code commenting for Codex completions, be used to check for implementation consistency and accuracy, and improve the quality of training data for Codex to name a few. Our approach of using comments with Codex should be reexamined under a variety of applications including program repair and defect prediction. This is an interesting future direction.



\section{Threats to Validity}

\subsection{Internal Validity}
\noindent
\textit{ManySStuBs4J}

We assume the samples in this dataset are mostly actual bugs. The authors report that changes related to refactoring, are removed, but some non-bug-fixing commits may remain. 

\hfill \break
\noindent
\textit{Manual Inspection}

We use automated matching to determine whether the models produce a known SStuB or patch. However, it is possible the automatic evaluation misses semantically equivalent but syntactically different bugs or patches. This could potentially hide the true number of bugs and patches. To reduce this threat to our results, we have three independent raters (the authors) inspect random samples from \textit{davinci-codex} completions (for the cases with no-comments, and the cases
with non-buggy comments) that matched \underline{neither} bug nor fix (we call
this unmatched subset ``dark matter''). 

With fair agreement, Fleiss Kappa 0.40, the independent raters found the vast majority (over 80\%) to be inappropriate code completions (neither bug nor fix --- just wrong), and sparsely little bugs or patches for the no-comment case with Davinci~\autoref{tab:sstubrate}. With moderate agreement, Fleiss Kappa 0.6, the independent raters found a smaller majority (about 70\%) of inappropriate code generations in the ``dark matter sample'' for  the non-buggy comment case, \autoref{tab:sstubratecomments}. The independent ratings all found that, even in the
``dark matter'' sample, adding comments in the prompt resulted
in substantial increase in patches. While we acknowledge that Codex non-match completions pose a threat to our findings, our sampled examination of this ``dark matter'' in both settings (with and without comments) suggests that adding comments does help LLMs avoid the generation of SStuBs.

\hfill\break
\noindent
\textit{Data Leakage from LLM Training}

We cannot independently verify that the ManySStuBs4J dataset is excluded in the training of the models since none of the models' training data is published. ``Data leakage'' is traditionally a concern when evaluating the performance of language models as data seen during training might artificially inflate results. In our case, data leakage will bias the model towards an outcome either the bug or the fix. We examined the latest fix date for the studied SStuBs and found 100\% of the SStuBs were fixed by February 2019 \textit{and} the earliest data collected for training is 2020 (\textit{cushman}) and 2021 (\textit{davinci}). If there is data leakage from ManySStuBs4J, we postulate the models would most likely see the \textit{fixed} version of the code, but intriguingly, the models still produce 2x more SStuBs! 




\hfill\break
\noindent
\textit{Reproductions of Generations}

Depending on hyper-parameters, Codex models are nondeterministic in their text generation (for the same prompt). 
We sampled the top-1 completion for each ManySStuBs4J sample across all models (Codex, PolyCoder, and CodeGen),  with and without comments. 

\subsection{External Validity}
\noindent
\textit{ManySStuBs4J}

Generalizability is subject to the limits of ManySStuBs4J. The dataset consists of Java single statement bugs; our results  may not generalize to other languages, or less simple bugs. PySStuBs\cite{kamienski2021pysstubs} and TSSB-3M\cite{richter2022tssb} are larger, and cite different SStuB patterns. The ManySStuBs4J dataset is the appropriate size given our constraints on available compute, and also API access to Codex. 
We were limited by OpenAI's rate ceiling of 20 requests per minute; on local hardware, the largest model, CodeGen 16B takes over a day for a run with a single prompt on ManySStuBs4J. 

\hfill \break
\noindent
\textit{Models at Scale}

Language models are getting ever larger. Results may vary with the next generation of models. 

\section{Conclusion}
Most importantly, we find that Codex and other large language models \emph{significantly help avoid  human-produced simple, stupid bugs}! In our best case, around 30\% (5149) SStuBs were actually patched (avoided) by Codex Davinci.
Still, we find that large language models might produce many more SStuBs than patches. \emph{First},  Codex and PolyCoder produce nearly twice as many SStuBs as fixes. \emph{Second}, and very worryingly, Codex generated SStuBs apparently took significantly longer to resolve and that the SStuBs appear to be as natural as the correct statements. Our results show that AI pair programming can introduce SStuBs, and the manner in which developers are known to use such tools is not conducive to avoiding SStuBs. Still, though the models were somewhat SStuB prone, even out-of-the-box LLMs could have avoided as many as 2,300 SStuBs had developers used code completion instead of writing them. 

Since the simple, stupid bugs are quite obvious after detection, we explore the idea of guiding AI assistants by adding comments describing high-level functionality. The proposed strategy of communicating functional intent to Codex with comments improved the bug/patch ratio substantially. Finally, we explore minor misunderstandings in the intended functionality by using buggy comments and find that Codex may not require strict correctness in comments to avoid SStuBs. Our results suggest that good commenting practices, even in an automatic setting, can help other developers and Codex, especially in an era where AI generated code is regularly committed. 

Overall, our findings are somewhat promising, LLMs may help avoid \emph{at least some} simple, stupid bugs!

\bibliographystyle{IEEEtran} \balance 
\bibliography{references}

\end{document}